%% file: main.tex
\documentclass[UKenglish,cleveref, autoref,pdfa]{lipics-v2021}
\pdfoutput=1

\input{preamble}

\nolinenumbers 

\begin{document}

\title{AI Writes Code, Humans Pay the Debt.\\An Empirical Study on the Sustainability and Evolution of Agent-Generated Code}
\titlerunning{AI Writes Code, Humans Pay the Debt}

\author{Antonino Coppola}{University of Southern Denmark, Vejle, Denmark}{antonino@imada.sdu.dk}{XXXX}{}
\author{Matteo Esposito}{University of Oulu, Finland}{matteo.esposito@oulu.fi}{https://orcid.org/0000-0002-8451-3668}{}
\author{Rick Kazman}{University of Hawaii, USA}{kazman@hawaii.edu}{https://orcid.org/0000-0003-0392-2783}{}
\author{Valentina Lenarduzzi}{University of Southern Denmark, Vejle, Denmark}{lenarduzzi@imada.sdu.dk}{https://orcid.org/0000-0003-0511-5133}{}

\authorrunning{A.Coppola, M. Esposito, R. Kazman, and V. Lenarduzzi}

\Copyright{A.Coppola, M. Esposito, R. Kazman, and V. Lenarduzzi}

\category{Registered Report}


\maketitle

\begin{abstract}
  \textbf{Context.} The increasing adoption of Generative AI coding agents in software engineering is transforming how developers implement and maintain code. While these systems provide short-term productivity benefits, their long-term impact on software quality and technical debt remains unclear.  

\noindent\textbf{Aim.} We aim to investigate how agent-generated code affects software quality, focusing on issue localization accuracy, the introduction of technical debt, and its evolution over time.

\noindent\textbf{Method.} We will conduct a large-scale mining software repositories study using the SQuaD dataset, employing a candidate set of 628k issue tickets. We will generate agent-based implementations for these issues, and compare them with the actual commits done by developers using static analysis metrics and tools. We will analyze differences at the commit level and across releases, and we will consider multiple LLM-based Agents selected through a systematic benchmarking strategy.

\noindent\textbf{Expected Results.} We expect to provide empirical evidence on the trade-offs introduced by agent-based development, including differences in localization accuracy, variations in technical debt introduction, and potential divergence in long-term evolution. We expect the results to highlight variability across LLMs, to enrich our understanding of software evolution with Agents, and to inform more responsible adoption of Agents in software development.
\end{abstract}

\maketitle

\section{Introduction}
\label{sec:Intro}

\input{Section/Introduction.tex}

\section{Empirical Study Design}
\label{sec:CS}
\input{Section/CaseStudy.tex}

\section{Threats to Validity}
\label{sec:Threat}
\input{Section/Threats.tex}

\section{Related Work}
\label{sec:relworks}
\input{Section/RelatedWork.tex}
\section{Conclusion}
\label{sec:Conclusions}
\input{Section/Conclusion.tex}
\subparagraph*{Declaration on the use of generative AI}
The authors used ChatGPT for suggestions on improving textual clarity. All research design, data analysis, interpretations, and manuscripts were created by the authors themselves.

\bibliography{main}
\end{document}

%% file: preamble.tex
\usepackage{algorithmic}

\usepackage{graphicx}

\usepackage{multirow}

\usepackage{xurl}
\usepackage{enumitem}

\usepackage[table,xcdraw]{xcolor}
\usepackage[many]{tcolorbox}

\usepackage{hyperref}

 \usepackage{todonotes}
\definecolor{main}{HTML}{CFCFCF}  
\definecolor{sub}{HTML}{CFCFCF}   

\tcbset{
    sharp corners,           
    colback = white,         
    before skip = 0.2cm,     
    after skip = 0.5cm       
}

\definecolor{boxbg}{gray}{0.92}      
\definecolor{boxborder}{gray}{0.55}  

\definecolor{main}{HTML}{CFCFCF}  
\definecolor{sub}{HTML}{CFCFCF}   

\usepackage[framemethod=tikz]{mdframed}
\usepackage{etoolbox} 
\usepackage{todonotes}
\newmdenv[
  skipabove=0.1\baselineskip,
  skipbelow=0.1\baselineskip,
  linewidth=0.5pt,
  linecolor=boxborder,
  backgroundcolor=boxbg,
  roundcorner=6pt,
  innerleftmargin=10pt,
  innerrightmargin=10pt,
  innertopmargin=8pt,
  innerbottommargin=8pt
]{boxC}

\newcommand{\boxCTitle}[1]{
  \noindent\textbf{#1}\par\vspace{4pt}
}

\newcounter{keyTakeAwaysCounter}

\newcounter{keyRQAnswerCounter}

\newcounter{RQCounter}

\newcounter{keyLimitationsCounter}

%% file: Section/Introduction.tex
The rapid adoption of Large Language Model (LLM)-based Generative AI coding agents is reshaping how developers write, modify, and maintain software. These tools can already generate syntactically correct and functionally plausible code for tasks such as bug fixing, feature implementation, and refactoring, becoming part of everyday development workflows. However, despite clear short-term productivity gains, their long-term impact on software quality and maintainability remains largely unexplored~\cite{adnan2026aiagentsgeneratemicroservices,fawareh2026significance}.

Software sustainability is commonly examined through Technical Debt (TD), which captures the trade-off between short-term gains and long-term quality, including suboptimal design decisions and maintainability issues~\cite{ernst2021, LenarduzziSLR2021}. If unmanaged, TD increases maintenance costs, limits system evolution, and reduces developer productivity~\cite{ernst2021,robredo2025evaluating}. In this context, a key question emerges: do coding agents introduce new forms of TD or amplify existing ones?

There are two concerns that arise from the use of coding agents. First, AI-generated code may differ from human-written code~\cite{patel2024comparative,cotroneo2025human-written}. Even when correct, it can be more verbose, less consistent, or misaligned with project conventions~\cite{adnan2026aiagentsgeneratemicroservices}. While these issues may not cause immediate defects, they can degrade maintainability and contribute to TD over time. However, there are no empirical studies that comprehensively investigate these concerns.
Second, it is unclear whether agents can correctly identify the locations of maintainability issues within a system. This task requires structural understanding and contextual reasoning~\cite{falessi_enhancing_2023,carka_effort-aware_2022}. While developers rely on experience, poor modularization choices by agents may lead to unnecessary or unnecessarily complex changes in the future, further increasing TD.

To address these gaps, our registered report aims to conduct a large-scale mining study on the impact of agent-generated code on software quality and TD. In our study, we will reconstruct historical issue-fixing contexts, revert repositories to pre-fix states, generate agent-based fixes, and compare them with human-written solutions using static analysis tools.
Therefore, we expect our study to provide the following \textbf{contributions}:

\begin{itemize}
\item A large-scale evaluation of how accurately coding agents localize issues within a codebase.
\item A comparison between agent-generated and real-world human-written code in terms of software quality and TD.
\item An analysis of how software quality and TD evolve over time under agent-driven changes.
\end{itemize}

%% file: Section/CaseStudy.tex
This section presents the empirical study design according to Wohlin et al. guideline~\cite{wohlin_experimentation_2024}.
We will provide all the raw data, all scripts, model prompts, and analysis configurations in a FAIR\footnote{\url{https://www.go-fair.org/fair-principles/}} replication package on Zenodo.

\begin{figure*}
    \centering
    \includegraphics[width=0.9\linewidth]{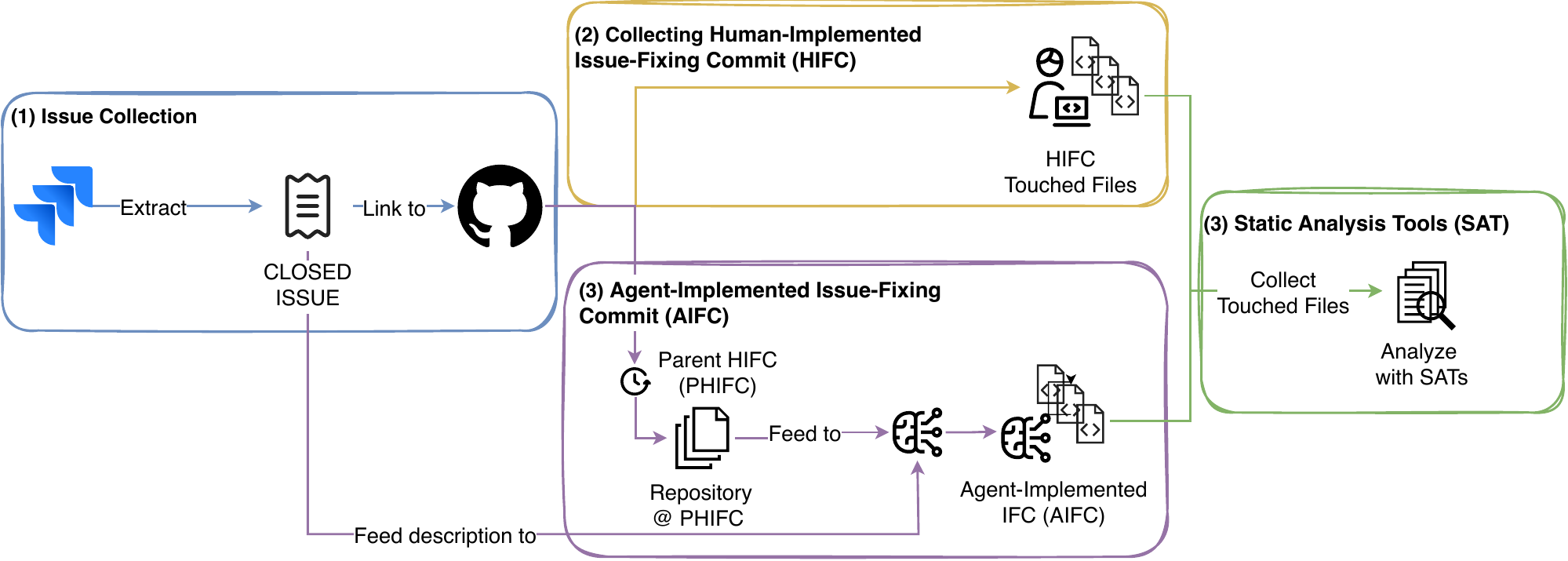}
    \caption{Data Collection Workflow}
    \label{fig:datacollection}
\end{figure*}

\subsection{Goal and Research Questions}
\label{sec:RQs}
We aim to analyze the code produced by agents during issue implementation tasks to evaluate their ability to understand, implement, and sustain software changes, from the viewpoint of TD and maintainability, relative to human developers.  

Therefore, we derived the following research questions (RQs) from our goal:

\begin{boxC}
\textbf{RQ$_1$.} To what extent can agents accurately determine the software entities that must be modified to address a reported issue?
\begin{itemize}[
    align=left,
    leftmargin=1.5em,
    labelsep=0.5em,
    itemsep=0.3em,      
    topsep=0.3em,      
    parsep=0pt,         
    partopsep=0pt    
]
    \item[\textbf{RQ$_{1.1}$}] Can agents accurately identify the \textbf{package} in which the modification should be made?
    \item[\textbf{RQ$_{1.2}$}] Can agents accurately determine which \textbf{class} contains the code that must be modified?
    \item[\textbf{RQ$_{1.3}$}] Can agents accurately locate the specific \textbf{method} that needs to be changed to resolve the issue?
\end{itemize}
\end{boxC}

Understanding where an issue should be resolved within a codebase is a central step in software maintenance and evolution. Issue localization is traditionally performed by developers who rely on domain knowledge, project history, and tool support to determine which files or software entities need to be modified~\cite{carka_effort-aware_2022}. With the rise of agents as assistants in software engineering tasks, an open question concerns their ability to perform this localization reliably. 
More specifically, a change is defined as the modification, addition, or removal of a software entity touched by the implementation commit. For method-level analysis, we rely on the entities identified through the extracted diffs and static analysis tooling, allowing us to consistently map both human- and agent-generated modifications to the corresponding software entities. 
In fact, we are keen on investigating whether agents can identify the package ($RQ_{1.1}$), the class ($RQ_{1.2}$), and the specific method ($RQ_{1.3}$) that should be touched, or created, according to the human-implemented reference fix. 
However, we acknowledge that the human-implemented fix represents a pragmatic rather than absolute ground truth for localization. In some cases, an agent may modify different software entities while still producing a valid, maintainable, or even superior solution, i.e., outcome-neutral alternatives. Therefore, in addition to measuring accuracy against human implementation, we will analyze a statistically significant stratified sample of divergent cases to determine whether they represent genuine localization errors or \textit{outcome-neutral alternative implementations}. Since manually inspecting all cases would not be feasible at the scale of this study, we will adopt a stratified sampling strategy to preserve representativeness across projects, issue types, and software entities, to avoid treating alternative implementation choices as necessarily inferior. 
We operationalize localization performance using standard information-retrieval metrics, including accuracy, precision, recall, F1-score, and MCC, computed by comparing agent-touched entities against the human-implemented reference fix. Since localization requires increasingly fine-grained contextual understanding from packages to classes and methods, we expect agents to perform differently across software entity granularities. Therefore, we test the following hypotheses:

    \begin{itemize}[
    align=left,
    leftmargin=3em,
    labelsep=0.5em,
    itemsep=0.3em,      
    topsep=0.3em,      
    parsep=0pt,         
    partopsep=0pt    
]
    \item[$H_{01}$] There is no significant difference in agent localization performance across package-, class-, and method-level software entities.
    \item[$H_{11}$] Agent localization performance significantly differs across package-, class-, and method-level software entities, with lower performance expected at finer-grained entities.
\end{itemize}

However, accurately identifying where to intervene does not guarantee high-quality fixes. Even when the correct entity is located, changes may still introduce technical debt or degrade code quality. Hence, we ask:

\begin{boxC}
\textbf{RQ$_2$.} Do agent-generated commits introduce more TD than human ones?
\end{boxC}

Agents may indeed produce semantically and syntactically valid code, yet the long-term quality of such code remains uncertain. Evidence suggests that AI-generated code may be more verbose or less maintainable~\cite{takerngsaksiri2025code,wang2025beyond}. Evaluating whether agent-written commits introduce more TD clarifies their sustainability and engineering maturity.
We measure TD and maintainability metrics using static analysis tools, including technical debt, maintainability, modularity, code smells, code health, cyclomatic complexity, duplication rate, and static rule violations (see Section~\ref{sec:context}). We investigate two complementary aspects: first, whether TD and maintainability profiles differ among LLM-based agents; and second, whether agent-generated fixes differ from human-generated fixes. Hence, we test the following hypotheses:

    \begin{itemize}[
    align=left,
    leftmargin=3em,
    labelsep=0.5em,
    itemsep=0.3em,      
    topsep=0.3em,      
    parsep=0pt,         
    partopsep=0pt    
]
    \item[$H_{02a}$] There is no significant difference in TD-related indicators and maintainability metrics among fixes generated by different LLM-based agents.
    
    \item[$H_{12a}$] Fixes generated by different LLM-based agents exhibit significantly different TD and maintainability profiles.
    
    \item[$H_{02b}$] There is no significant difference in TD-related indicators and maintainability metrics between agent-generated fixes and human-generated fixes.
    
    \item[$H_{12b}$] Agent-generated fixes exhibit significantly different TD and maintainability profiles from human-generated fixes, with higher TD-related indicators and lower maintainability expected for agent-generated fixes.
\end{itemize}

In fact, recent findings suggest that LLMs tend to generate overly coupled and unnecessarily complex solutions even for relatively simple problems~\cite{zhu2026aigeneratedsmellsanalysiscode}, thereby supporting the expected directionality of our hypotheses.
All in all, TD and code quality evolve as software grows; therefore, differences observed at the commit level may compound over time. Hence, we ask:

\begin{boxC}
\textbf{RQ$_3$.} Does TD accumulate faster in agent-generated code over software releases?
\end{boxC}

TD compounds over time, commit by commit~\cite{Cunningham1992}. If agents introduce, amplify, or mitigate debt differently from humans, this might affect the long-term sustainability of a project as it evolves. By re-implementing historical commits, we can observe whether cumulative debt and maintainability trends diverge between human-driven and agent-driven software evolution trajectories.
We operationalize TD accumulation over time as the release-level evolution rate of TD-related indicators and maintainability metrics. In particular, we compare the slopes of human-driven and agent-driven trajectories across releases, where steeper positive slopes for TD-related indicators indicate faster debt accumulation, and negative slopes for maintainability-oriented metrics indicate stronger quality degradation. Therefore, we test the following hypotheses:

    \begin{itemize}[
    align=left,
    leftmargin=3em,
    labelsep=0.5em,
    itemsep=0.3em,      
    topsep=0.3em,      
    parsep=0pt,         
    partopsep=0pt    
]
    \item[$H_{03}$] There is no significant difference in the release-level evolution rate of TD-related indicators and maintainability metrics between human-driven and agent-driven software evolution trajectories.
    
    \item[$H_{13}$] Agent-driven software evolution trajectories exhibit significantly different release-level evolution rates from human-driven trajectories, with steeper TD growth and stronger maintainability degradation expected across releases.
\end{itemize}

\subsection{Study Context and LLM Selection}
\label{sec:context}
We base \textbf{project selection} on SQuaD, introduced by Robredo et al.~\cite{robredo2025squad}, a multi-dimensional, time-aware dataset of software quality metrics from 450 mature open-source projects (e.g., Apache, Mozilla, FFmpeg, Linux kernel). It integrates nine state-of-the-art static analysis tools, namely SonarQube, CodeScene, PMD, Understand, CK, JaSoMe, RefactoringMiner, RefactoringMiner++, and PyRef, unifying over 700 metrics at the method, class, file, and project levels. Covering 63,586 releases, SQuaD also includes version control, issue tracking, vulnerability data (CVE/CWE), and process metrics for JIT defect prediction and over 628k issues. 
Although the schema contains 725 raw attributes extracted from CK, JASOME, SonarQube, Understand, PMD, and CodeScene, many metrics are duplicated or semantically equivalent across tools. Therefore, we performed a consolidation process that merged equivalent metrics, such as coupling, size, and classic OO metrics (e.g., WMC, DIT, LCOM), into canonical representations, i.e., with the same name. After deduplication, 349 static-analysis columns were reduced to 230 canonical metrics (34.1\% reduction). We narrowed down the tool selection to three complementary and industrially adopted analyzers, namely SonarQube, Understand, and CodeScene, that multi-language static analysis tools. Since the focus of this study is maintainability and technical debt, we filtered the metric space to retain only metrics mapped to maintainability-related quality characteristics as defined in ISO 25010, ISO 9126, and ISO 5055. In particular, we retained metrics related to maintainability, analyzability, modifiability, stability, complexity, coupling, cohesion, duplication, inheritance, encapsulation, coverage, and maintainability-oriented rule violations, while excluding metrics primarily targeting security, reliability, or performance; resulting in a final pool of \textbf{110 maintainability-oriented} metrics used throughout the study.
We implement an \textbf{agent} using the OpenCode framework\footnote{\url{https://opencode.ai/}} to integrate LLMs into software engineering tasks. We consolidated three independent LLM ranking sources accessed in March 2026: Artificial Analysis\footnote{https://artificialanalysis.ai}, LLMstats\footnote{https://llm-stats.com}, and OpenCompass\footnote{https://rank.opencompass.org.cn}, to identify the most performant models for code-related tasks across different evaluation metrics. Artificial Analysis employs Terminal Bench Hard and SciCode benchmarks, LLMstats leverages the Code-Arena benchmark, and OpenCompass utilizes the LiveCodeBench metric for assessment. Following consolidation, we computed the median ranking position across these sources and subsequently selected the top-performing models demonstrating consensus across multiple ranking systems. Based on this methodology, we identified three models for experimental evaluation: GLM-5, Qwen3.5-397b, and Kimi K2.5.

\subsection{Data Collection}
\label{sec:DataCollection}

We divided our data collection workflow into three processing steps (Figure~\ref{fig:datacollection}).

\textbf{(1) Issue Collection.}  
We extract closed issues from each project's issue-tracking system (e.g., Jira or GitHub Issues). We select closed issues because they represent completed tasks with verified outcomes and links to the commits that resolved them. For each issue, we collect metadata such as title, description, labels, timestamps, and references to implementation commits. We exclude issues without explicit links to resolving commits to ensure clear traceability between issues and code changes. To avoid contamination from AI-generated contributions, we limit the dataset to commits dated before January $1^{st}$, 2021.

\textbf{(2) Collecting Human-Implemented Issue-Fixing Commits.}  
We link each issue to its corresponding implementation commit(s), referred to as Human-Implemented Issue-Fixing Commits (HIFC). Since issue-to-commit linkage is inherently noisy, we adopt a \textbf{best-effort approach} based on regular expression matching used in our previous study~\cite{robredo2025analyzing}. We use: \textit{JIRA pattern:} \texttt{[A-Z]{2,}-\textbackslash d+}; \textit{GitHub pattern:} \texttt{\#\textbackslash d+} ;\textit{Numeric pattern:} \texttt{\textbackslash d{4,}}.  
We acknowledge that this approach may miss some issue-fixing commits due to heterogeneous commit practices, and we discuss this as a threat to validity (Section~\ref{sec:Threat}). For each HIFC, we extract commit metadata, including message, diff, touched files, author, timestamp, and parent commit. We retrieve the parent commit and reconstruct the repository's pre-implementation state, which represents the exact context developers had before the fix.

\textbf{(3) agent-based re-implementation.}  
We provide the agent with the issue description and the corresponding codebase snapshot, and we instruct it to produce a patch that resolves the issue. Thus, obtaining the agent-Implemented Issue-Fixing Commit (AIFC).  \textbf{We will not provide any information from the human implementation, such as diffs, commit messages, future code states, nor access to tools, i.e., git, that would reveal the commit history to the agent.}

\textbf{(4) Static Analysis.}
We transform HIFC- and AIFC-touched files into standardized patches, validate their syntactic correctness, normalize non-semantic differences, and analyze the repositories using Static Analysis Tools (SATs).

More specifically, \textbf{syntactic correctness} is enforced using tree-sitter-java, applied symmetrically to both human and agent-generated code. For each modified file, we recover the same pre-fix snapshot, parse it, and map changed lines to AST entities. Using the same process for both sides ensures that disagreements reflect real localization differences rather than tooling artifacts. Files that fail parsing are excluded. We selected tree-sitter because it handles partial code well and provides stable ASTs across Java 8–21.
\textbf{Non-semantic differences} are normalized by comparing Fully Qualified Names (FQNs), e.g., org.apache.commons.numbers.fraction.Fraction\#pow, rather than patch text. This makes comparisons invariant to formatting, comments, declaration order, variable names, file paths, and import ordering. We do not normalize semantic equivalences (e.g., \textit{for} vs \textit{while}), as our focus is on which entity the agent identifies rather than how the fix is implemented.

We will leverage the pipeline proposed in~\cite{robredo2025squad} to extract all the measurable metrics in our context, including code quality metrics and TD measures from tools such as SonarQube and CodeScene (Table~\ref{tab:sqa-tools}). We will also expand the SQuaD dataset with software architecture anti-patterns and modularity scores over time using DV8~\cite{div8}.
Finally, for each issue, we construct a dataset that includes: the issue description and metadata; the human-implemented commit; the agent-generated commit; and the static analysis results for both implementations and software entities at granular levels.
\input{Tables/metrics_nocitations}

\subsection{Data Analysis}
\label{sec:DataAnalysis}

This section outlines how we will analyze the data to answer the research questions and evaluate the associated hypotheses.
Throughout the data analysis, to identify the correct statistical techniques to test the relevant hypothesis, we must assess the distribution of the collected data~\cite{esposito_critical_2026}. Hence, we always pose the following hypothesis over the collected data:

\begin{itemize}[
    align=left,
    leftmargin=3em,
    labelsep=0.5em,
    itemsep=0.3em,      
    topsep=0.3em,      
    parsep=0pt,         
    partopsep=0pt    
]
    \item[H$_{0\mathcal{N}}$] \textit{The collected data is normally distributed.}
    \item[H$_{1\mathcal{N}}$] \textit{The collected data is not normally distributed.}
\end{itemize}

We test \emph{\textbf{H$_{0\mathcal{N}}$}} using Anderson-Darling (AD) test \cite{anderson1952asymptotic}. AD tests whether data points are sampled from a specific probability distribution, in this case, a normal distribution. According to Mishra et al., \cite{mishra2019descriptive}, the Shapiro-Wilk (SW) test \cite{shapiro1965analysis} would be more appropriate when using smaller datasets with fewer than 50 samples. Given the large sample size, we employ Anderson-Darling, which is particularly sensitive to departures from normality, especially in distribution tails~\cite{stephens1974edf}. Furthermore, we also highlight that across all RQs, we report the statistical test, $p$-value, effect size, and a practical interpretation of the results. 
We use a significance level of $\alpha = 0.05$. Omnibus tests are evaluated at this threshold. For post-hoc pairwise comparisons, we apply Bonferroni correction, using an adjusted significance threshold $\alpha' = \alpha/k$, where $k$ denotes the number of pairwise comparisons~\cite{dunn1961,wohlin_experimentation_2024}. To mitigate the trade-off between Type~I and Type~II errors and avoid relying on a single correction strategy, we additionally report Benjamini--Hochberg-adjusted (BH) $p$-values as a sensitivity analysis~\cite{Benjamini1995FDR}. Similar to Robredo et al.~\cite{robredo_what_2026}, we analyze both the pairs that remain significant \textbf{under the intersection} of the two correction strategies and those that are significant \textbf{under only one adjustment method}, i.e., borderline findings. In fact, Bonferroni strongly controls false positives but may hide meaningful effects, whereas Benjamini–Hochberg retains higher statistical power at the cost of a less conservative correction. Reporting both allows us to distinguish highly robust findings from weaker but potentially meaningful effects.

\textbf{Evaluating agent-Based Issue Localization Accuracy (RQ$_1$).}
We test $H_{01}$ by analyzing differences in localization performance across software entity granularities, namely, packages, classes, and methods. More specifically, we compare the distributions of the localization metrics (Accuracy, Precision, Recall, F1-score, and MCC) across granularity levels.
 
Before selecting the statistical test, we assess the normality of the metric distributions. Since localization performance is evaluated on the same issues and agent configurations across different software entity granularities, the study follows a repeated-measures design. If normality is not satisfied, we apply the \textit{Friedman test}, followed by post-hoc \textit{Wilcoxon signed-rank tests} with multiple-comparison correction. Otherwise, we apply a \textit{repeated-measures ANOVA}, followed by paired-sample $t$-tests with correction.
We additionally analyze whether localization performance decreases at finer-grained entities (from packages to classes and methods). To quantify effect sizes, we report Kendall’s~$W$ for Friedman analyses, rank-biserial correlation ($r_{rb}$) for non-parametric post-hoc comparisons, and partial $\eta^2$ for repeated-measures ANOVA analyses.

\textbf{Assessing TD Introduced by agent-generated Code (RQ$_2$).}
We compare the TD profile of agent-generated code against that of human-written commits. We rely on the metrics extracted in Step~4 of the data collection process, which include maintainability debt, code smells, code health, cyclomatic complexity, duplication rate, and static rule violations. Following Ernst et al.~\cite{ernst2021}, we interpret these metrics as proxies for maintainability and design debt.
For each issue, we analyze the full codebase after applying the human-generated patch and, separately, each agent-generated patch for different LLM-based configurations. We then execute the static analysis pipeline on the resulting repository states and compare the extracted metric values and violations. By repeating this process for each issue-resolving commit, we construct parallel observations for human-generated and agent-generated implementations.

We test $H_{02a}$ by analyzing differences in TD-related and maintainability metrics across LLM-based agents. Since the same issues are evaluated across multiple configurations, the study follows a repeated-measures design. After assessing normality, we apply either the \textit{Friedman test} with post-hoc \textit{Wilcoxon signed-rank tests}, or a \textit{repeated-measures ANOVA} with paired-sample $t$-tests, both with multiple-comparison correction.
We test $H_{02b}$ by comparing human-generated and agent-generated implementations across all LLM configurations. As the same issues are evaluated under all conditions, we follow the same repeated-measures procedure adopted for $H_{02a}$.
To quantify effect sizes, we report Kendall’s~$W$ for Friedman analyses, rank-biserial correlation ($r_{rb}$) for non-parametric post-hoc comparisons, partial $\eta^2$ for repeated-measures ANOVA, and Cohen’s $d$ for paired parametric comparisons.

\textbf{Analyzing the Evolution of TD in Agent-Driven Systems (RQ$_3$).}
For each project and each metric, we obtain paired slopes $(\beta_1^{Human}, \beta_1^{Agent})$, representing the release-level evolution rates derived from two parallel full-repository trajectories: (i) the observed human-driven trajectory and (ii) the counterfactual agent-driven trajectory obtained by injecting the aggregated repository-level deltas into the observed release metrics.  Since both trajectories are derived from the same project, metric, and release sequence, the resulting observations are paired and follow a repeated-measures design.
We test $H_{03}$ by comparing the distributions of the extracted slopes across projects. Before selecting the statistical procedure, we assess the normality of the slope distributions. If normality cannot be assumed, we apply the \textit{Wilcoxon signed-rank test}, as it represents the non-parametric alternative for comparing two related groups. Otherwise, we use the paired $t$-test as the parametric alternative.
Given our expectation that agent-driven trajectories may exhibit steeper TD growth and stronger maintainability degradation, we additionally analyze whether the observed differences follow the expected directionality across releases. To quantify the magnitude of the observed differences, we report the rank-biserial correlation ($r_{rb}$) for nonparametric paired comparisons and Cohen's $d$ for parametric paired comparisons.
As a sensitivity analysis, we repeat the slope-based comparison separately for each agent--LLM configuration to verify whether the observed effects remain consistent across different models.
Furthermore, when sufficient release-level observations are available, we employ linear mixed-effects models as a robustness analysis. Since multiple releases are observed within the same project, release-level observations are not independent and may follow project-specific temporal trends. Therefore, we model each metric as a function of release index, implementation type (human vs.\ agent), and their interaction, while including project-level random intercepts:
\[
m_{p,r} = \beta_0 + \beta_1 \cdot r + \beta_2 \cdot \text{Type} + \beta_3 \cdot (r \times \text{Type}) + u_p + \epsilon
\]

where $u_p$ captures project-level variability. The interaction term $(r \times \text{Type})$ captures whether the evolution rates differ between human-driven and agent-driven trajectories. We use this model as a robustness check to verify whether the slope-based results remain consistent when considering all release-level observations simultaneously.

\textbf{Contingency Plan.} We will adopt a sampling-based strategy when hardware or time constraints prevent exploring all collected issues~\cite{robredo_what_2026}. We will select a representative subset stratified across projects, issue types, and granularity levels, prioritizing traceable issues and maintaining balance across projects and releases. We will report the procedure transparently and perform sensitivity analyses when feasible to ensure robust results and mitigate validity threats. 
Given the computational cost and execution time associated with large-scale agent-based issue re-implementation, sampling may be employed as part of the study design rather than solely as a fallback contingency. The adopted sampling strategy depends on the research objective. For RQ$_1$ and RQ$_2$, we employ a stratified issue-level sampling strategy to preserve representativeness across projects, issue types, releases, and software entities. For RQ$_3$, where preserving temporal continuity is essential to analyze software evolution, we prioritize repository-level sampling and analyze most, or all when feasible, issue-fixing commits within selected repositories. Consequently, the interpretation and generalizability of the results will be contextualized according to the adopted sampling strategy.

%% file: Tables/metrics_nocitations.tex
\begin{table}[tb]
\centering
\footnotesize
\caption{Overview of SATs included in SQuaD~\cite{robredo2025squad}.}
\begin{tabular}{p{1.5cm}p{1.3cm}p{10cm}}
\hline 
\textbf{Tool} & \textbf{\#Metric} & \textbf{Aspect covered} \\ 
\hline 
CK & 88 & Calculates class-level and method-level code metrics in Java projects. \\
JaSoMe & 70 & Mines file, package, class \& method quality metrics in Java projects. \\ 
Understand & 111 & Mines file, class \& entity quality metrics for multiple languages. \\ 
SQ & 192 & Calculates several quality metrics \& verifies the code’s compliance against a specific set of ``coding rules''. \\ 
PMD & 114 & Runs coding rules against source files to find violations. \\ 
CodeScene & 22 & Computes per-file comprehensive code health checks. \\ 
\hline 
\end{tabular}
\label{tab:sqa-tools}

\end{table}

%% file: Section/Threats.tex
This section discusses the threats to validity, including internal, external, construct validity, and reliability~\cite{wohlin_experimentation_2024}. 

\textbf{Construct Validity.} 
We rely on static analysis tools (e.g., SonarQube, CodeScene, and Understand) and software quality metrics as proxies for TD and maintainability. Although these metrics cannot fully capture the multidimensional nature of TD, they are widely adopted in empirical software engineering research, and the SQuaD dataset was specifically designed to aggregate and harmonize validated metrics across large-scale repositories. Nevertheless, these metrics remain approximations rather than direct measurements of developers’ perceived debt. To mitigate this limitation, we rely on multiple complementary metrics and focus on consistent trends rather than isolated variations~\cite{Lefeveretal2021}.
We also acknowledge a limitation regarding coding assistants. In our study, we fix the orchestration framework (OpenCode) and vary only the underlying LLMs. Therefore, the findings should be interpreted as differences across LLMs within a fixed orchestration environment rather than a comparison of coding assistant frameworks in general. Consequently, the results may not generalize to alternative systems such as Claude Code, SWE-agent, or OpenHands, which may differ in prompting strategies, tool integration, or execution workflows. Exploring alternative designs that fix the LLM while varying the orchestration framework is left for future work.
Our metric normalization strategy may also introduce bias if normalization distorts results. To mitigate this, we apply metric-specific normalization rules. Moreover, differences between human- and agent-generated code may reflect stylistic variations rather than actual TD, so we focus on structural and maintainability-related indicators.

\textbf{Internal Validity.} We link issues to commits using regular expressions, which may introduce noise due to inconsistent commit messages. We mitigate this by relying on established patterns and selecting cases with clear traceability, although some noise may remain. Agent-generated outputs may also vary because of prompt design and model behavior. To reduce this variability, we standardize prompts and perform sensitivity analyses with multiple LLM-based agents.
Confounding variables may influence differences between human- and agent-generated implementations. In particular, issue quality and clarity may affect an agent’s ability to localize and implement fixes, while developer expertise and coding practices may affect the quality of baseline commits. Thus, some differences may reflect issue complexity or developer expertise rather than only differences between human- and agent-generated code. 
Our counterfactual trajectories approximate system evolution by injecting per-issue metric differences into release-level data. This assumes additive and independent changes, which may not capture interactions, path dependency, or long-term architectural effects. Moreover, the simulation remains anchored to the historically observed human-driven issue sequence, potentially introducing structural bias. Therefore, the proposed trajectories should be interpreted as a first-order approximation rather than a causal reconstruction of software evolution.

\textbf{External Validity.} Our findings are based on open-source projects and may not fully generalize to industrial settings. We include projects with diverse characteristics (e.g., size, activity, ecosystem) to broaden the scope. Using pre-2021 data avoids contamination from AI-generated code but may not reflect current practices.

\textbf{Conclusion Validity.} Since this is a Stage 1 Registered Report we do not report threats to conclusion validity.

%% file: Section/RelatedWork.tex
Due to space constraints, we will curate a broader, up-to-date body of related work to ensure comprehensive coverage of the research landscape at the time this study is implemented. At this stage, however, and to the best of our knowledge, the studies discussed below represent the closest empirical works to our focus on the maintainability of AI-generated code.

Recent empirical investigations have examined the quality of AI-generated code. For instance, Cotroneo et al.~\cite{cotroneo2025human-written} conducted a large-scale comparison between human-written and AI-generated code, revealing differences in defects, vulnerabilities, and complexity. Similarly, Wang et al.~\cite{wang2025beyond} investigated coding style inconsistencies introduced by large language models, highlighting issues that can hinder readability and comprehension. Takerngsaksiri et al.~\cite{takerngsaksiri2025code} reinforced this perspective through an industrial case study, showing that readability remains a critical concern in real-world settings. Earlier work by Patel et al.~\cite{patel2024comparative} also reported structural differences between AI-generated and human-written code, suggesting potential downstream implications for maintenance activities.
Overall, the current state of the art provides an important baseline, indicating that AI-generated code cannot be evaluated solely in terms of functional correctness or short-term productivity. At the same time, existing studies primarily focus on static quality indicators and short-term characteristics, without explicitly examining how AI-generated code contributes to the introduction and accumulation of TD over time. Consequently, the long-term maintainability implications of AI-generated code remain insufficiently understood. Therefore, understanding how AI-generated code influences TD accumulation and software evolution across releases remains an open research challenge.

%% file: Section/Conclusion.tex
We investigate the impact of agent-generated code on software quality and technical debt through a controlled and reproducible empirical study. We compare human-written and agent-generated implementations across multiple projects and metrics to build a comprehensive understanding of how LLM-based Agents influence software maintenance practices.
We aim to provide empirical evidence on the strengths and limitations of agent-based code generation, focusing on issue localization, the introduction of technical debt, and long-term evolution. We analyze not only whether differences emerge, but also how these differences appear across software entities, quality dimensions, and over time.
We also contribute a methodological framework for evaluating AI-generated code in realistic development scenarios. Our approach combines counterfactual trajectories, multi-metric analysis, and sensitivity analysis across different LLM configurations. We expect this framework to support future research on the sustainability and reliability of AI-assisted software engineering.
We expect our findings to inform both researchers and practitioners about the trade-offs involved in adopting LLM-based agents. We aim to support more informed decisions on how to integrate these technologies into software development workflows while preserving software quality and long-term sustainability.